\documentclass[12pt,a4]{article}

\usepackage{amsmath}
\usepackage{amssymb}
\usepackage{graphicx}
\usepackage{color}

\newcommand{\rf}[1]{(\ref{#1})}
\newcommand{\beq}{\begin{equation}}
\newcommand{\eeq}{\end{equation}}
\newcommand{\bea}{\begin{eqnarray}}
\newcommand{\eea}{\end{eqnarray}}

\newcommand{\e}{{\rm e}}

\newcommand{\lam}{\lambda}
\newcommand{\Lam}{\Lambda}

\newcommand{\al}{\alpha}

\newcommand{\Om}{\Omega}

\newcommand{\Del}{\Delta}

\newcommand{\kp}{\kappa}

\newcommand{\oh}{\frac{1}{2}}

\newcommand{\prt}{\partial}

\newcommand{\cH}{{\cal H}}

\newcommand{\tp}{\hat{p}}

\def\void{}
\def\labelmark{}

\newenvironment{formula}[1]{\def\labelname{#1}
\ifx\void\labelname\def\junk{\begin{displaymath}}
\else\def\junk{\begin{equation}\label{\labelname}}\fi\junk}%
{\ifx\void\labelname\def\junk{\end{displaymath}}
\else\def\junk{\end{equation}}\fi\junk\labelmark\def\labelname{}}

{\ifx\void\labelname\def\junk{\end{array}\end{displaymath}}
\else\def\junk{\end{array}\right.\end{equation}}
\fi\junk\labelmark\def\labelname{}\def\junk{}
}

\newcommand{\beqv}{\begin{formula}{}}

\vspace{12pt}

\newcommand{\Gs}{G_{\rm s}}
\newcommand{\tF}{\tilde{F}}

\begin{document}

\rightline{\today}

\begin{center}
\vspace{24pt}
{ \Large \bf Can a  late-time cosmological model  based on 
baby universe absorption explain the z-variation of  w?}

\vspace{24pt}

{\sl J.\ Ambj\o rn}$\,^{a,b}$,
and {\sl Y.\ Watabiki}$\,^{c}$

\vspace{10pt}

{\small

$^a$~The Niels Bohr Institute, Copenhagen University\\
Blegdamsvej 17, DK-2100 Copenhagen \O , Denmark.\\
email: ambjorn@nbi.dk
\vspace{10pt}

$^b$~Institute for Mathematics, Astrophysics and Particle Physics
(IMAPP)\\ Radbaud University Nijmegen, Heyendaalseweg 135, 6525 AJ, \\
Nijmegen, The Netherlands

\vspace{10pt}

$^c$~Dept. of Physics, High Energy Theory Group,\\
Institute of Science Tokyo\\
2-12-1 Oh-okayama, Meguro-ku, Tokyo 152-8551, Japan\\
{email: watabiki@th.phys.titech.ac.jp}

}

\end{center}

\vspace{24pt}

\begin{center}
{\bf Abstract}
\end{center}

\noindent

We point out that a simple late-time cosmological model where our Universe 
can absorb  "baby universes"explains the exponential expansion of our universe
without the need of a cosmological constant and leads to a
 z-dependence of the parameter w(z)  in the equation of state. 
In this model  w(z)  is less than -1 for z sufficiently large.

\newpage

\section{Introduction}\label{intro}

If one assumes that the late time cosmology is well described by the FLRW metric
 \beq\label{j3}
 ds^2 = -N^2(t)dt^2 + a^2(t) d\Om_d,\qquad d \Om_d = \sum_{i=1}^d dx_i^2,
 \eeq
 the time 
evolution can be derived from the minisuperspace action
\beq\label{j2}
 S =  \int dt  \,\Big(-\frac{3}{8} \frac{\dot{v}^2}{N v} -  N v( \lam + \kp \rho(v) )\Big).
 \eeq
 Units used are $c=\hbar = 1$, and we use 
 \beq\label{j3a} 
 v(t) = \frac{1}{\kp} a^3(t),\quad \kp = 8\pi G, \quad \mbox{$G=$ the gravitational constant,}
 \eeq
  as variable rather than the scale factor $a(t)$. $v(t)$ is proportional 
 to the $d$-dimensional 
 spatial volume at time $t$ and below we will just call it the spatial volume. 
 $\rho (v)$ denotes the matter density.  
 \beq\label{n3}
 \kp \rho(v) = \kp \rho_{\rm m}(v) + \kp \rho_{\rm r} (v) = 
 3 H_0^2 \left( \Om_{\rm m} \frac{v_0}{v} + \Om_{\rm r} \left(\frac{v_0}{v}\right)^{4/3}\right),
 \eeq
 where $v_0$ denotes the spatial volume at the present time $t_0$ and $H_0$ the Hubble 
 constant, i.e.\ the Hubble parameter $H(t) = \dot{a}(t)/a(t) =  \dot{v}(t)/3 v(t)$ at 
 $t_0$. Finally $\lam$ denotes the cosmological constant.
 
If $p$ denotes the momentum conjugate to $v$, the Hamiltonian related to the action $S$ given 
by eq.\ \rf{j2} will be 
\beq\label{j30}
 \cH = N {v} \,\Big( -\frac{3}{4} p^2+ \lam + \kp \rho(v) \,\Big).
\eeq
The Friedmann equation associated with \rf{j30} is 
\beq\label{j30a}
\frac{3}{4} p^2 = \lam +\kp \rho(v) ,
\eeq
and when $t \leq t_{\rm LS}$, the time of last scattering, we can ignore $\lam$. 
Let $T(t)$ denote the temperature of the CMB at times $t \geq t_{\rm LS}$. Then 
$a(t_{\rm LS})/a(t_0) = T(t_0)/T(t_{\rm LS})$ and since we know  
$T(t_{\rm LS})$  from atomic physics and can measure $T(t_0)$ and $H_0$, we can write
\beq\label{j30b}
\frac{3}{4} p^2(t_{\rm LS}) =  \left( \frac{\rho_{\rm m}(v_0)}{(1+z(t_{\rm LS}))^3} + 
\frac{\rho_{\rm r}(v_0)}{(1+z(t_{\rm LS}))^4}\right) ,
\eeq
where $z(t) = a(t_0)/a(t) -1$ is the redshift at time $t$.
Thus, $v (t_{\rm LS})/v_0$ is known and 
$p(t_{\rm LS})$ is known if $\rho_{\rm m}(v_0)$ and $\rho_{\rm r}(v_0)$ 
are known, and finally $t_{\rm LS}$ can be found by integrating \rf{j30a} 
starting from  $t=0$, $z(0) = \infty$, to $z(t_{\rm LS})$. 
We just mention these textbook facts since we will use
 them in a model with a modified Friedmann equation.

In \cite{aw1,aw2} we proposed  modifications of the standard minispace Hamiltonian 
\rf{j30} that allow our universe to absorb other universes during its time evolution\footnote{To describe the emission instead of absorption of baby universes, one simply changes the sign of $g$
to be negative. The mathematics of the two cases are almost identical and in this paper we focus on the case $g > 0$ and refer to baby universe absorption.}.
In the modified models one could freely choose the distribution of spatial volumes of the 
absorbed universes, and we considered two limits: one where the spatial volume of an absorbed 
universe was infinitesimal, and one where there is a certain self-consistency between 
our universe and the universes absorbed. In this article we will concentrate on the latter 
model and below we will discuss  what ``self-consistency" means.

Since \cite{aw1} was published the new DESI results seem to favor an equation \cite{desidr2}
of state 
\beq\label{n4}
P(z) = w(z) \;\rho(z),   
\eeq
where $\rho(z)$ and $P(z)$ are the  energy density and the pressure of dark energy,
and where $w$ is $z$-dependent and  $w < -1$ for $z$ sufficiently large. Already 
in \cite{aw1}  we noted that  such a behavior is a 
natural consequence for the  kind of model  we consider. We here 
take the opportunity to check to what extent  our model can accommodate the latest DESI results.

\section{The late-time cosmological model}

In our model the standard minisuperspace Hamiltonian \rf{j30} is replaced by
\bea
\cH &=& N  {v}\Big(  - \,\frac{3}{4} \big(p^2 -2g \tilde{F}( p)\big) +\lam+ \kp \rho(v)\Big)\label{j38a}\\
&=& N {v} \Big( - f(p) + \kp \rho (v) \Big), \quad  f(p) =-\frac{3}{4} (p +\al)  
\sqrt{ ( p-\al)^2 +\frac{2 g}{\al} } 
\label{j38}
\eea
where $\al$ satisfies 
\beq\label{j7}
\quad \al^3 -  \frac{4}{3} \lam\al - g = 0.
\eeq
A new coupling constant $g$ is introduced and it is assigned to the process where two 
universes merge to one universe (see \cite{aw1} and references therein for a detailed discussion). 
Eq.\ \rf{j7} is a consistency condition that ensures that in the case where $\rho (v) =0$, 
for  given coupling constants $g, \lam$, 
the Hartle-Hawking wave functions of the universes involved 
in the mergers are identical.  In \cite{aw1} we denoted the model with this $f(p)$
the gcdt-model, since the equations \rf{j38} and \rf{j7} was first derived in a 2d quantum
gravity model called generalized causal dynamical triangulations \cite{al1}. 
The function $\tF(p)$ has an expansion in inverse power of $p$ and in the early universe 
where $v(t)$ is small and thus $\rho(v)$ and $|p|$ are large, $\tF(p)$ (and $\lam$) will not be important. 

The equations of motion  for  the Hamiltonian \rf{j38} are 
\beq\label{j41}
\frac{\dot{v}}{v} = -f'( p), \quad \dot{ p} =  f( p) + \frac{1}{3} \rho_{\rm r}(v), 
\quad f (p) = \kp \rho_{\rm m}(v) + \kp \rho_{\rm r} (v),  
\eeq
where the last equation is the ``Friedmann'' equation that follows from $\prt \cH/\prt N  =0$.
 Assuming that the universe expands at large $t$ it follows that   $p < -\al$, and  
 from the Friedmann equation it is seen that  $p(t) \to -\al$ for large $t$. 
 Thus that the expansion is exponential for large $t$:
 \beq\label{n5}
 v(t) \propto \e^{-f'(-\al) t}, \quad f'(-\al) = -\frac{3}{4} \sqrt{ 4\al^2 + \frac{2g}{\al}} = 
 - \sqrt{ 3 \lam + \frac{27 g}{8 \al}}.
 \eeq
 For $g =0$ we have the usual exponential expansion of the de Sitter universe, and increasing 
 $g$ will only increase this expansion in accordance with the intuition that our Universe in the model
 absorbs the spatial volume of other universes. In fact one can even take $\lam =0$ and still
 have an exponential expanding universe at late time, the expansion only dictated by the absorption
 of other universes and the rate being given by 
 \beq\label{n6}
 v(t) \propto \e^{ \sqrt{\frac{27}{8}} \;g^{1/3} t}.
 \eeq
In the following we will assume $\lam =0$ in our generalized minisuperspace model, 
such that we have exchanged the coupling constant 
$\lam$ with a new coupling constant $g$ that controls the strength of the merging of universes.

 For the standard $\Lam$CDM 
 model the value of $\lam$ is such that it plays (almost) no role for $t \leq t_{\rm LS}$, the time 
 of last scattering, as already mentioned. 
 The same will be the case for $g$ in our generalized model. The new coupling 
 constant $g$ will be so small that we can ignore it for $t \leq t_{\rm LS}$, and thus we will 
 have the same time evolution of $v(t)$ as for the  $\Lam$CDM model for  $t \leq t_{\rm LS}$.
 Given $\kp \rho_{\rm m} (v_0)$ and  $\kp \rho_{\rm r}(v_0)$ we can solve \rf{j41} for 
 $v(t)/v_0$ and $p(t)$, starting out at $t_{\rm LS}$ where we know the values, and integrating 
 the equations to $v(t)/v_0 =1 $ that will then determine our present time $t_0$. We thus have 
 two models, the $\Lam$CDM model and our multiverse model, that start out at 
 $t_{\rm LS}$ with the same values of  $v(t_{\rm LS})/v_0$ and $p(t_{\rm LS})$, 
 and thus the same value $H(t_{\rm LS})$ of the Hubble parameter, but when integrated
 to the region $t \gg t_{\rm LS}$, where $\lam$ and $g$, respectively,  are important, will 
 behave differently. In particular the time $t_0$ will be different in the two models and 
 the Hubble parameter $H(t)$ will be different, as will the Hubble constant $H_0= H(t_0)$.
 
 In \cite{aw1,aw2} we made a simple analysis of this difference, by just taking the 
 value of $\rho(v)$ that was  available from the best fits to the $\Lam$CDM model. 
 In principle $\rho(v)$ is determined by physics at times $t < t_{\rm LS}$ and is then 
 the same for the $\Lam$CDM model and our  model.  By insisting that the locally measured 
 $H_0 \approx 72.6$ km/s/Mpc (see \cite{h01}) is the correct value of $H_0$, the value $g$
 was fixed. We noticed then that the  $H(z)$ calculated for this value of $g$ agreed well with the 
  low $z$ measurements of $H(z)$. In addition, for the same value of $g$,
 we could calculate the parameter $S_8$, measuring the large-scale structure clustering of matter, 
 to be $S_8 \approx 0.76$, also in agreement with local observations.

 The best fit to all data, {\it excluding the local measurement of $H_0$}, 
 using the $\Lam$CDM model, leads to a value
 of $\lam$ such that  $H_0 \approx 67$ km/s/Mpc and $S_8 \approx 0.83$, resulting in  
 the so-called $H_0$ and $S_8$ tensions with local measurements.
 In \cite{aw1,aw2} we formulated it as follows: assuming the local measurements of $H_0$ 
 are correct our model seems to describe late time cosmology better than the $\Lam$CDM
 model. In \cite{cline1} a more careful analysis of the our model was performed, 
 using the same fitting as for the $\Lam$CDM model. The result was that the $\Lam$CDM model 
 fitted the data (again excluding local measurements of $H_0$) 
 somewhat better\footnote{\label{ftn1} In \cite{cline1} 
 one found $\chi^2 = 4780.4$ for the $\Lam$CDM model and   $\chi^2 = 4783.5$ for the 
 multiverse model, i.e. a $\Del \chi^2 = 3.1$. This implies (according to standard AIC) 
 that the probability that 
 the multiverse model should be preferred rather than the  $\Lam$CDM model, 
 based on the present data, is 
 $\e^{-\Del \chi^2/2} = 0.21$. }. 
 However, even if we have not performed the analysis, it is relatively clear from  the results 
 \cite{cline1} that {\it if} one was fitting with the additional constraint that  
 $H_0\approx$ 72.6 km/s/Mpc 
 (i.e.\ assuming that the local measurements give the correct value of $H_0$), our
  model would be doing as well as  the $\Lam$CDM model. 
 With this situation in mind, we now turn to the equation of state parameter $w(z)$.

\section{ A z-dependent  w}

Using \rf{j41} we can  write the generalized Friedmann equation $f(p) = \kp \rho(v)$ 
as follows
\beq\label{def3}
\Big(\frac{\dot{a}(t)}{a(t)}\Big)^2 = \frac{\kp \rho(v)}{3} + \frac{\kp \rho_{\rm f}(v)}{3},
\quad
\kp \rho_{\rm f} ( p) = \frac{1}{3} \big(f'( p)\big)^2 - f( p),
\eeq
where we can view $ \rho_{\rm f} ( p) $ as the energy density associated to the 
merging of other universes with our Universe. If we define the 
corresponding  pressure $P_{\rm f}$ by the energy conservation equation 
\beq\label{def6}
\frac{d}{dt} ( v \rho_{\rm f}) +  P_{\rm f} \frac{d}{dt} v = 0,
\eeq
we obtain  the {\it formal equation of state parameter} $w_{\rm f}$:
\beq\label{def8}
w_{\rm f}  = \frac{P_{\rm f}}{\rho_{\rm f}} = 
\frac{ f(p)   \Big( \frac{2}{3} f''( p) -1\Big)}{ \frac{1}{3} \big(f'( p)\big)^2 - f( p)} -1.
\eeq
When $t \gg t_{\rm LS}$ we can ignore $\rho_{\rm r}(v(t))$ and the only contribution to $w$ 
comes from $\rho_{\rm f}$. As a function of $t$ or $z(t)$, $w_{\rm f}(z)$ 
interpolates between $w_{\rm f} = -1.5$ at $z \to \infty$, i.e.\
$t \to 0$, and $w_{\rm f} = -1$ for $z \to -1$, i.e.\ $t \to \infty$ (where the universe expands exponentially). We now want to compare this behavior with the $w(z)$ extracted from the DESI data
\cite{desidr1}, \cite{desidr2}.

\begin{figure}[t]
\centerline{\scalebox{0.25}{\rotatebox{0}{\includegraphics{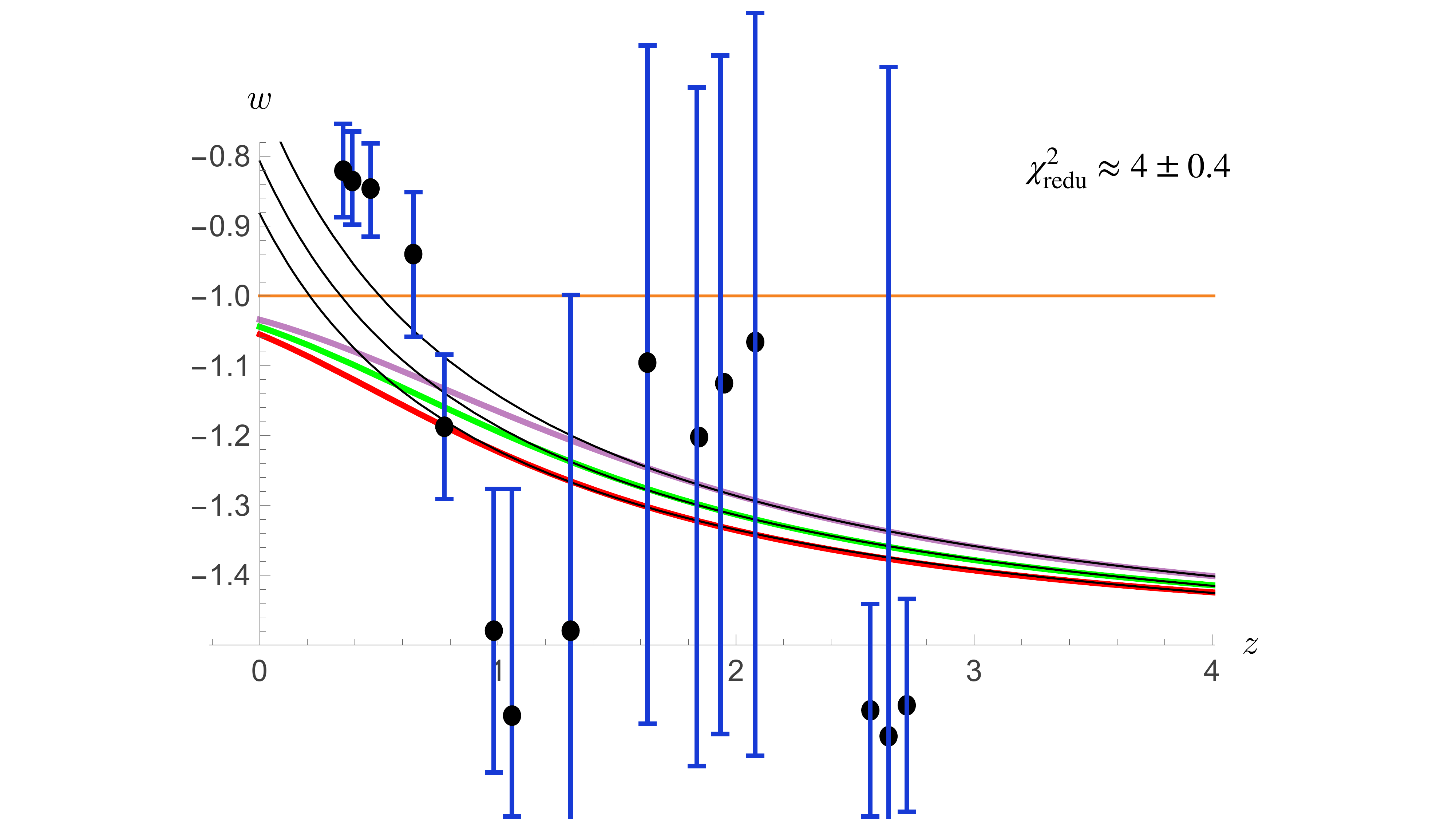}}}}
\caption[fig1]{{\small
The data points and error bars are imported from \cite{desidr2}, Fig.\ 7.
 The red curve is  $w_{\rm f}(z)$ for $H_0 = 72.6$. The reduced $\chi^2$ value comparing 
 the curve with data results in a  
 $\chi^2_{\rm redu} \approx 4\pm 0.4$.
 A corresponding  $w(z)$ (black curve) given by 
 \rf{n7} with  $(w_0,w_a) = (-0.866, -0.695)$ is a best fit to $w_{\rm f}(z)$ for $z > 0.8$. 
 The green curve is $w_{\rm f}(z)$ with the first order 
 correction in $G_{\rm s}$ included in $f$ (eq. \rf{a1}), $G_{\rm s}=-4$, and the corresponding 
 $w(z)$ (black curve) is given by 
 \rf{n7} with  $(w_0,w_a) = (-0.789, -0.778)$. The purple curve is  $w_{\rm f}(z)$ for 
 $G_{\rm s} = -8$, calculated using a Pad\'e[2/1] approximant as described in the Appendix. 
 The corresponding $w(z)$ (black curve) has   $(w_0,w_a) = (-0.690, -0.884)$.
 }}
\label{fig1}
\end{figure}

The  DESI DR2 data and analysis \cite{desidr2} 
indicate that a $z$-dependent $w(z)$ in the equation of state
\rf{n4} leads to a better fit to the late time data than the $w=-1$ 
predicted by the $\Lam$CDM model. In addition a number of the parameterizations of $w(z)$ 
suggest that $w(z) < -1$ for $z > 0.5$. Values of $w(z) < -1$ have been difficult to explain 
in  physical motivated extensions of the $\Lam$CDM model 
(see \cite{cline2,cline3} for recent discussions), and most of the parameterizations
of $w(z)$ used in fits are chosen for their simplicity and are not related to any physics.
Here we will just concentrate on the commonly used $w_0,w_a$ parameterization 
\beq\label{n7}
w(z) = w_0 + \frac{z}{z+1} w_a .
\eeq
 Fig.\ \ref{fig1} shows $w_{\rm f}(z)$ (the red curve), 
 calculated using the value $H_0 = 72.6$. In addition 
a best fit of the form \rf{n7} to this $w_{\rm f}(z)$ is shown. 
The fit is performed for $z \in [0.8,4]$. It is clear that 
the functional form \rf{n7} cannot fit $w_{\rm f}(z)$ both for small values of $z \in [0,0.8]$ and 
for  $z \in [0.8,4]$. The fit corresponds to the values $(w_0,w_a) = (-0.87,-0.70)$.

The best values of $w_0,w_a$ extracted from the data, i.e. DESI data and other data set 
depend on the combination of data set. As an example of best values, 
the combination DESI DR1, CMB  and pantheon+ data leads to \cite{desidr1}
\beq\label{n8}
w_0 =  -0.83 \pm 0.06, \qquad w_a =  -0.75 \pm 0.26,
\eeq
compatible with the values used in Fig.\ \ref{fig1}. However, clearly the data used 
in  Fig.\ \ref{fig1}, which uses the Union3 data  rather than pantheon+ data would lead to 
somewhat  different $(w_0,w_a)$ values. 

From Fig. \ref{fig1} we have calculated the reduced $\chi^2$ values of our 
function $w_{\rm f}(z)$ compared to the data points and error-bars and we obtain  that 
a rough estimate leads to 
\beq\label{n8a}
\chi^2_{\rm redu} \approx 4\pm 0.4 .
\eeq
Thus we are far away from a perfect fit. It is clear that it is the small $z$ data points that
do not agree well with $w_{\rm f}(z)$.

\section{Inclusion of topology changes}

\begin{figure}[t]
\centerline{\scalebox{0.25}{\rotatebox{0}{\includegraphics{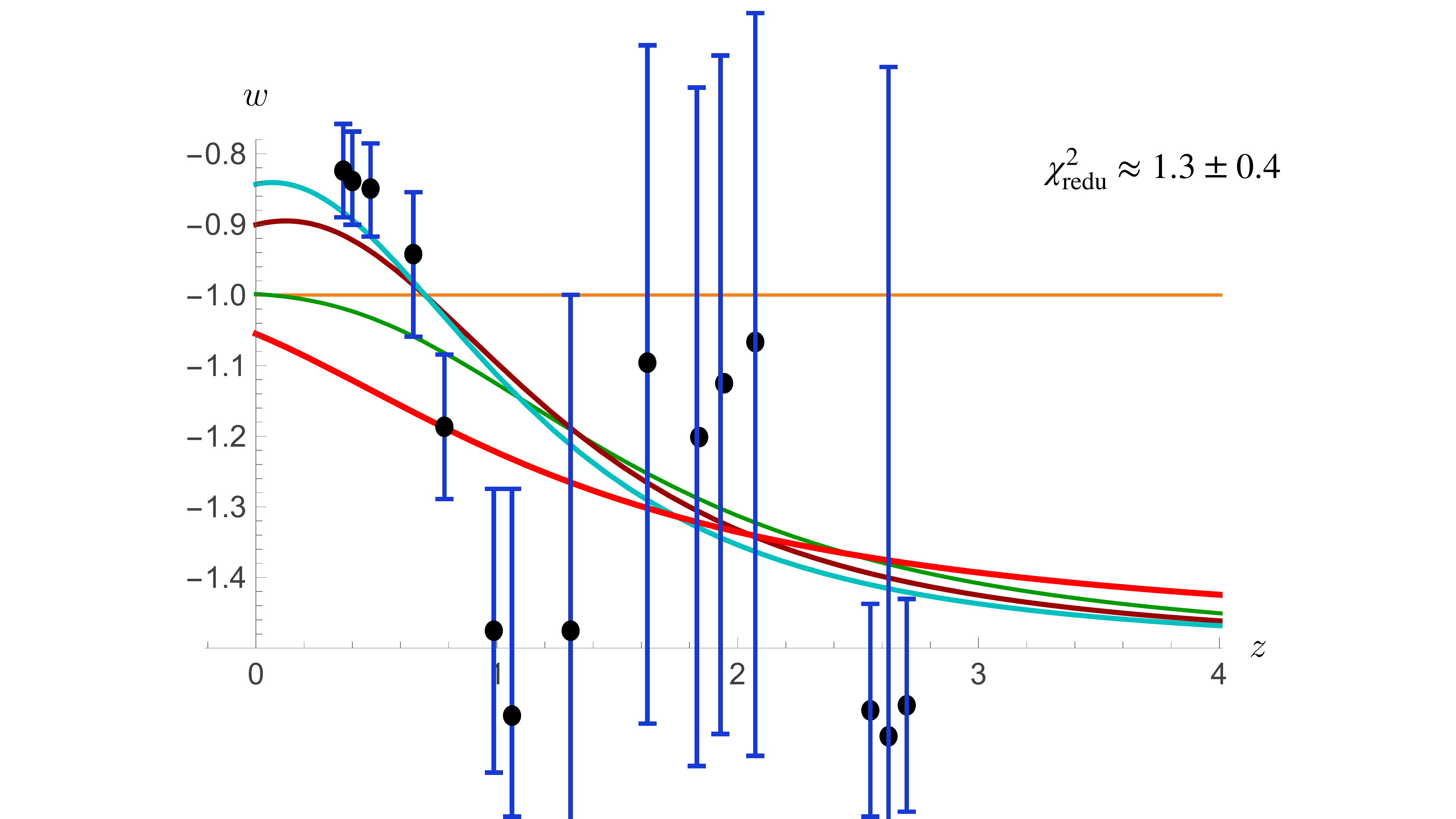}}}}
\caption[fig2]{{\small
 As in Fig.\ \ref{fig1} the data points are imported from Fig.\ 7 in \cite{desidr2}.
 The curve in red is the $w_{\rm f}(z)$ curve for $f(p)$ given by \rf{j38}, green curve is 
 $\tilde{F}_1$ from eq.\ \rf{m1a} with $a_1=3$, the dark red curve $\tilde{F}_2$ with $a_2=1$ and 
 the cyan curve $\tilde{F}_3$ with $a_3=0.4$. The $\chi^2_{\rm redu}$ refers to the 
 $\tilde{F}_3$
 curve.}}
\label{fig3}
\end{figure}

\begin{figure}[t]
\vspace{-1cm}
\centerline{\scalebox{0.15}{\rotatebox{0}{\includegraphics{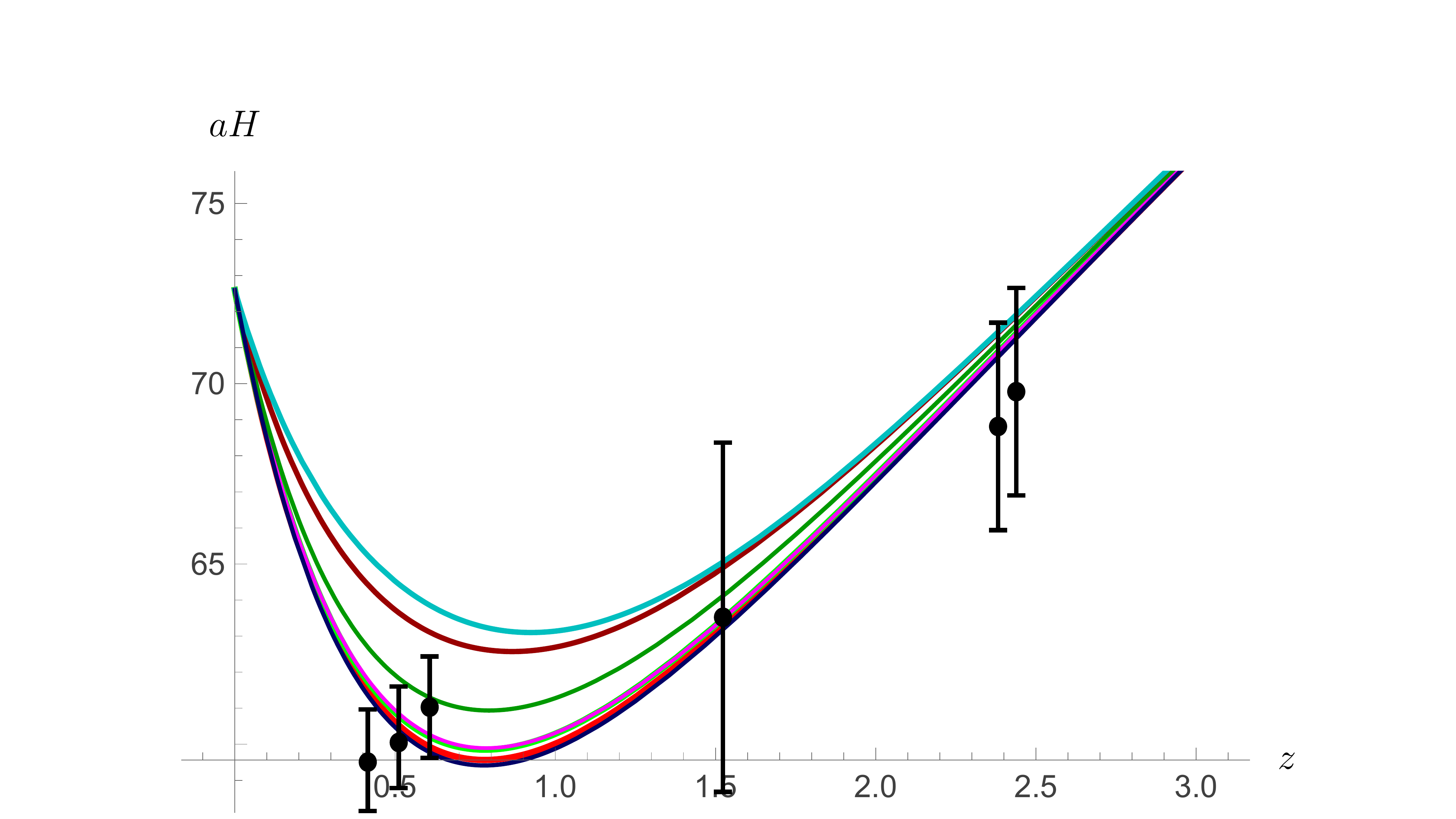}}}
\scalebox{0.6}{\rotatebox{0}{\includegraphics{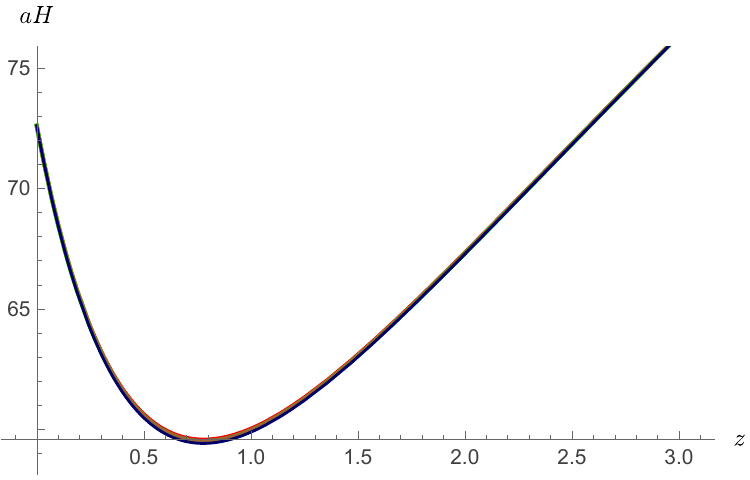}}}
\vspace{-2mm}}
\caption[fig4]{{\small
 The left panel shows $H(z)/(1+z)$  calculated for $f(p)$ given by \rf{m1} and \rf{m1a}
 (the three upper curves (dark green, dark red and cyan), using $\tF_1$, $\tF_2$  and $\tF_3$ with 
 $a_1=3$, $a_2=1$ and $a_3 =0.4$), 
 as well as $H(z)/(1+z)$ calculated 
 for $f(p)$ given by \rf{n10}-\rf{n10a} with $\Gs =-4$ and to 0th, 1st and 2rd orders in
 $\Gs$,  (red, green and magenta)
 Also, in blue, is  shown $H(z)/(1+z)$ calculated to all orders in $\Gs$ for $\Gs =1$. 
 The $H(z)$ data the same as used in \cite{aw1}. On the 
 right panel we show $H(z)/(1+z)$ again calculated for $f(p)$ given by \rf{n10}-\rf{n10a},
 but with $\Gs=1$. Same color labelling as for the lower 4 graphs on the left panel and in addition
 in brown is the calculation to 3rd order in $\Gs$.
 }}
\label{fig4}
\end{figure}

Can one improve the agreement between $w_{\rm f}(z)$ and the data for small $z$?  
As already remarked, our model is not unique. One can change the distribution of 
the spatial volume of the absorbed universes by hand. It is indeed possible 
to change the function $\tilde{F}(p)$ in such a way that  $w_{\rm f}(z) > -1$ for small $z$ while still being 
less than $ -1$ for larger $z$. In Fig.\ \ref{fig3} 
we have shown the result of using 
\beq\label{m1}
f(p) = \frac{3}{4} ( p^2 -2g \tilde{F}(p) ) 
\eeq
with the following choices of $\tF(p)$
\beq\label{m1a}
 \tilde{F}_1(p)\! =\! \frac{1}{\sqrt{p^2+ a_1 g^{2/3}}},\quad \tilde{F}_2(p) \!=\!  \frac{-p}{p^2+ a_2 g^{2/3}}, \quad
 \tilde{F}_3(p) \!=\! \frac{-p^3}{(p^2+ a_3 g^{2/3})^2}.
\eeq
The fits have an additional parameter $a_i$. We have not tried to adjust the 
parameter to a best fit to data, but it is clear from Fig.\ \ref{fig3} that both $a_2$ in $\tF_2(p)$ 
and $a_3$ in $\tF_3(p)$ can be adjust to provide a good  fit. For the $\tF_3(p)$ used in 
Fig.\ \ref{fig3} we have 
$\chi^2_{\rm redu} \approx 1.3\pm 0.4$. However, with this choice of parameters 
($a_2$ and $a_3$) the Hubble parameters $H_{f_2}(z)$ and $H_{f_3}(z)$ calculated 
using functions $f_2$ and$f_3$ from \rf{m1} and \rf{m1a} corresponding to $\tF_2$ and 
$\tF_3$ do not fit very well with the measured $H(z)$, as shown in Fig.\ \ref{fig4}.  
It might be that one can 
find a function $\tF$ leading to acceptable fits to both $H(z)$ and $w(z)$ data, but 
we have not tried to do that, the reason being that  one would like some 
physical motivation for a given choice of $\tF$. The choice \rf{j38} was motivated by
the idea that  all 
universes, including our ``own" Universe, have the same Hartle-Hawking wave 
function if $\rho(v) = 0$. The 
choice that the absorbed universes have a matter content and that the Hartle-Hawking 
wave functions of such universes should be identical to the Hartle-Hawking 
wave function of our Universe, including matter, is maybe even more ``natural''. 
Unfortunately, we do not know how to 
calculate the Hartle-Hawking wave function when matter is included, 
even in a minisuperspace approximation. 
 
However, it seems natural to try to include  
more complicated changes of spacetime topologies in the model. 
Until now we have considered only the process that two universes can merge to one universe.
One could include processes where the individual universes are allowed to change 
spacetime topologies, and one can calculate the simplest topology change corrections to the function $f(p)$ given
by \rf{j38} (see \cite{gcdt1,gcdt2}). 
This introduces a new coupling constant  $G_{\rm s}$  (a ``string" coupling constant)
\beq\label{n10}
f(p) =-\frac{3\, s^2}{4} \left( (\tp+1) \sqrt{(\tp -1)^2 +2} + 2s\, \tilde{F}(\tp, \Gs)\right),
\eeq
\beq\label{n10a}
s \tF(\tp,\Gs) = \sum_{h=1}^\infty \Gs^h\, s\tF^{(h)}(\tp),\quad s = g^{1/3}, \quad \tp = p/s.
\eeq
Here  $h$ denotes 
the ``genus'' of the spacetime\footnote{The notion of ``genus'' relates to the 
original calculation in two-dimensional spacetime. In that case it 
was literally an expansion in the number of handles, $h$, of the two-dimensional manifold.},
and $\Gs$ the coupling constant associated with the change of spacetime
topology from $h$ to $h+1$.
The functions $\tF^{(h)}(\tp)$, $h\geq 1$, fall
off like $1/\tp^2$ for large $|\tp|$ and they are then not important for large values of $z$, but 
they could potentially change the behavior for small values of $z$ and this is what we want. 
We have calculated $\tF^{(1)}$, $\tF^{(2)}$ and $\tF^{(3)}$ (see Appendix). The expansion \rf{n10a}
is only an asymptotic expansion, and since we are interested in the series in the region where 
$|\tp|$ is not much larger than 1 (small $z$), 
only the first two orders can be trusted unless we take $G_{\rm s}$ small. In Fig.\ \ref{fig4}
we show the results of calculating $H(z)$ to various orders of $\Gs$ for $\Gs =-4$ and 
$\Gs =1$. 
Rather surprisingly, even for  $\Gs$ as large as  $-4$,  
the first and second order corrections to $H(z)$ are quite small, but one sees a clear change 
when one includes the  third order corrections. Thus, for this value of $\Gs$, we can only  trust 
the expansion \rf{n10a}  to second order. However, one can include the third order correction 
by using instead the Pad\'e[2/1] approximant. 
That seems to give good results even for larger values
of $|\Gs|$ (see the Appendix).  For $\Gs =1$ one can 
actually calculate  $\tilde{F}(\tp,\Gs)$ to all orders in $\Gs$ \cite{az} (it is the Laplace 
transform of the Airy function). One can then check that the Pad\'e[2/1] approximant , using 
up to 3rd order in $\Gs$ works very well for this value of $\Gs$ and one can observe 
that for this value of $\Gs$ the function  $H(z)$ is 
not changed much compared  to $H(z)$ calculated for $\Gs=0$ (that gives 
a good fit to the $H(z)$ data \cite{aw1,aw2}).

While $H(z)$ only involves the first derivative of $\tF(\tp,\Gs)$ wrt $\tp$, 
$w_{\rm f}(z)$  involves the second derivative, and the change as a function of $\Gs$
could be larger. However, it is not the case. 
The result is shown in Fig.\ \ref{fig1} for $G_{\rm s} = -4$ using the 1st
order correction and for $G_{\rm s} =-8$ using the Pad\'e[2/1] approximant.
As shown in Fig.\ \ref{fig5} in the Appendix one can obtain a positive $w_{\rm f}(z)$ for small 
$z$ by taking $G_{\rm s}$ very large and negative, but  such a large value is 
not ``natural''. In addition, while such values of $G_{\rm s}$ might give a better fit 
to the $w(z)$ data, the corresponding $H(z)$ will not give a good fit to 
the $H(z)$ data, i.e.\ we have the same situation as 
with the functions $\tF_2$ and $\tF_3$ above.

\section{Extending the baby universe  model}

Until now we have considered  a model where the $\Lam$CDM model's
cosmological constant $\lam$  is replaced by the coupling constant $g$ describing the
merging of spatial universes. However, it is of course possible to enlarge  the model. 
The simplest extension of the model is to leave
both $\lam$ and $g$ in the expression  \rf{j38} for $f(p)$. This has been analyzed in \cite{cline1},
studying various combinations of $\lam$ and $g$ greater than or smaller than zero. In this case it is
indeed possible to find values of $\lam$ and $g$, such  that the $\chi^2$ value from
comparing the model with data is less than the corresponding $\chi^2$ for the best 
fit of the $\Lam$CDM model. This is not surprising since already the 
``pure'' model with $\lam =0$ was doing quite well (see footnote \ref{ftn1}).
However, these extended models do not really do better than the $\lam=0$ model when 
comparing the calculated $w_{\rm f}(z)$ and the $w(z)$ determined from data, and they 
come at the price of having two coupling constants instead of one.

Another extension of the model is  to combine it with quintessence and consider 
adding a scalar field $\phi$ coupled minimally to gravity (see \cite{quint} for a review).  
These models in general lead to a $w(z)$ that varies with $z$, but it is difficult to 
obtain a $w(z) < -1$. Thus a combination of our model and quintessence
might result in a $w(z) > -1$ for small $z$ and $w(z) < -1$ for larger $z$. It is 
possible that one can find such an extended model that can fit all data, 
but the drawback is again that it involves a number of new parameters.

\section{Discussion}

Our late-time cosmological  model  has replaced the 
cosmological constant $\lam$ in the $\Lam$CDM model with a new coupling constant 
$g$, and the value of $g$ is then determined by insisting that the calculated value 
of the Hubble parameter $H(t)$ at the present time $t_0$ agrees with the locally 
measured value $H_0 = 72.6$. This allows us to {\it calculate} $w_{\rm f}(z)$ and 
it has some qualitative  similarity with the  $w(z)$ one has extracted from the data: a $w(z)$
decreasing with increasing $z$ and being less than $-1$ for sufficient large $z$. For $z < 0.5$ the 
agreement with data is not good and a simple $\chi^2$-fit to the data leads
to $\chi_{\rm red} \approx 4 \pm 0.4$. 
The agreement for small $z$ could be improved a little by including spacetime topology
changing corrections, but comes at the expense of  the introduction of a new coupling constant.
Good agreement for small $z$ could be obtained
by changing the function $\tF$ as in \rf{m1}-\rf{m1a}. Unfortunately we lack presently physical 
motivation for making such a change. 

However, let us stress simplicity of the  original model. It has just one coupling 
constant $g$ related to late time cosmology, 
and it is fixed just fitting to local model-independent measurements of 
$H_0$ and $T_0$ (the present CMB temperature). For this value of $g$
there is good agreement with most low $z$ measurements\footnote{Of course the 
standard $\Lam$CDM model also has only one parameter, the cosmological constant $\lam$,
related to late time cosmology. However, fixing it simply from the local measurements
of $T_0$ and $H_0$ does not lead to good agreement with other local measurements.} 
and it leads to  an equation of state parameter $w_{\rm f}(z)$ with a $z$ dependence sharing qualitative 
features with the $w(z)$ extracted from data. It explains the late time exponential expansion
of the universe without the need a cosmological constant, and recent work suggests that
it might be possible to explain the smallness of the coupling  constant $g$ from an
underlying microscoping theory \cite{aw3}. 
It thus seems worth to try to develop this theory further, 
in particular to try to understand the merging of universes in the context of ordinary 
GR.

\section*{Appendix}

\begin{figure}[t]
\centerline{\scalebox{0.9}{\rotatebox{0}{\includegraphics{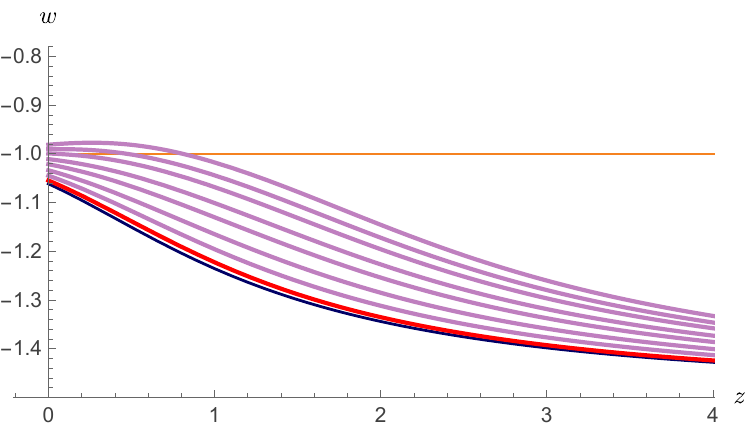}}}}
\caption[fig4]{{\small
 The calculated $w_{\rm f}$ for $\Gs = 1, 0, -4,-8,-12,-15,-18,-20, -22$ using the Pad\'e[2/1] approximant,
 the red curve being the 
 $\Gs =0$ curve. For $\Gs$ large negative $w_{\rm f}(z)$ becomes larger than $-1$ for 
 small values of $z$. }}
\label{fig5}
\end{figure}

The functions  $\tF^{(h)}(\tp)$ can be found by iteratively solving the Swinger-Dyson equations 
defining GCDT \cite{gcdt1}, but a better way is to use the results of topological recursion 
as described in \cite{wata}. At zeroth order we have ($\tp \leq -1$):
\beq\label{a0}
s \tF^{(0)}(\tp) := \oh(\tp+1) \sqrt{\tp^2 -2 \tp +3} +\oh \tp^2 =-\frac{1}{\tp} -\frac{3}{4} \frac{1}{\tp^2} + \cdots,
\eeq
the ``genus 0'' GCDT amplitude we have already used. 
The function $s \tF^{(1)}(\tp)$ can also be found  in \cite{gcdt1}:
\beq\label{a1}
s\tF^{(1)}(\tp) = \frac{(-\tp+ 3)(\tp^2 - 2\tp + 9)}{72(\tp^2 - 2\tp + 3)^{5/2}}, 
\eeq  
when adjusted for  sign changes and normalization, compared to \cite{gcdt1}   

The iteration of the Swinger-Dyson equations to next order leads to 
\beq\label{a2}
s\tF^{(2)}(\tp) =\frac{
\begin{pmatrix}         -17\tp^9 + 187\tp^8 - 1122\tp^7
        + 4522\tp^6 - 13319\tp^5+
      \\
         29073\tp^4 - 45252\tp^3
        + 41040 \tp^2 + 2430\tp - 23814
\end{pmatrix}}{15552(\tp^2 - 2\tp + 3)^{11/2}}.
\eeq
The expression for  $\tF^{(3)}(\tp)$ is quite long and is as follows
\beq\label{a10}
s \tilde{F}^{(3)}(\tp)
=
\frac{\begin{pmatrix}
         1684\tp^{15} - 28628\tp^{14}
        + 257652\tp^{13}
        -1577908\tp^{12} + 7282601\tp^{11}
      \\-
         26698041\tp^{10} + 80254609\tp^9
        - 201831221\tp^8 + 430169949\tp^7
      \\-
         782096733\tp^6 + 1210818861\tp^5
        - 1568258253\tp^4
      \\+
         1612434762\tp^3 -1150859178\tp^2
        + 424098666\tp -31847094
      \end{pmatrix}}
     {1679616
      (\tp^2 -2\tp + 3)^{17/2}}.
\eeq
The expansion $s \tF(\tp,\Gs)$ in powers of $\tp^{-n}$ and $\Gs^h$ is

\begin{align}\label{a3}
s \tilde{F}(\tp,\Gs)
\,=&\,\,
   \sum_{h=0}^\infty\Gs^h
     s \tilde{F}^{(h)}(\tp)
\nonumber
\\
=&\,\,
     s \tilde{F}^{(0)}(\tp)
   +
     \Gs
     s \tilde{F}^{(1)}(\tp)
   +
     \Gs^{2}
     s \tilde{F}^{(2)}(\tp)
   +
     \Gs^{3}
     s \tilde{F}^{(3)}(\tp)
   +
     \ldots
\nonumber\\
=&\,\,
  -\frac{1}{\tp} -\frac{3}{4\tp^2}
       +\frac{1}{\tp^4}
       +\frac{3}{2\tp^5}
       +\frac{9}{16\tp^6}
       - \frac{2}{\tp^7}
       + {\cal O}\Big(\frac{1}{\tp^8}\Big)
\nonumber\\
&\,\,
   -\Gs \Big(
         -\frac{1}{72\tp^2}
       + \frac{1}{36\tp^5}
       +\frac{25}{48\tp^6}
       + \frac{2}{\tp^7}
       + {\cal O}\Big(\frac{1}{\tp^8}\Big)
     \Big)
\nonumber\\
&\,\,
   - \Gs^2\Big(
         -\frac{17}{15552\tp^2}
       + \frac{17}{7776\tp^5}
       +\frac{53}{3456\tp^6}
       + {\cal O}\Big(\frac{1}{\tp^8}\Big)
     \Big)
\nonumber\\
&\,\,
   - \Gs^3\Big(
         \frac{421}{419904\tp^2}
       - \frac{421}{209952\tp^5}
       -\frac{1075}{559872\tp^6}
       + {\cal O}\Big(\frac{1}{\tp^8}\Big)
     \Big)
   + \ldots
\end{align}
or, as an expansion in $\tp^{-n}$:
\begin{align}\label{a4}
s \tilde{F}(\tp,\Gs)
=&\,\, -\frac{1}{\tp}
 -\Big(\frac{3}{4}
    - \frac{\Gs}{72}
    - \frac{17\Gs^2}{15552}
    +\frac{421\Gs^3}{419904}
    +{\cal O}(\Gs^4)
   \Big) \frac{1}{\tp^2}
 +
   \frac{1}{\tp^4}
\nonumber\\
&\,\,
 -\Big(-
      \frac{3}{2}
    + \frac{\Gs}{36}
    + \frac{17\Gs^2}{7776}
    - \frac{421\Gs^3}{209952}
    + {\cal O}(\Gs^4)
   \Big) \frac{1}{\tp^5}
\nonumber\\
&\,\,
 +\Big(
      \frac{9}{16}
    - \frac{25\Gs}{48}
    - \frac{53\Gs^2}{3456}
    + \frac{1075\Gs^3}{559872}
    + {\cal O}(\Gs^4)
   \Big) \frac{1}{\tp^6}
\nonumber\\
&\,\,
 - \big(
      2
    + 2\Gs
    + {\cal O}(\Gs^4)
   \big) \frac{1}{\tp^7}
 + {\cal O}\Big(\frac{1}{\tp^8}\Big).
\end{align}

From these expansions one might think that $|\Gs| < 1$ is needed in order to trust the expansion
to third order in $\Gs$. However, we find that for $|\Gs| \leq 2$ there is little change in $H(z)$, 
$\Om_{\rm m}(z)$ and $w_{\rm f}(z)$ when the higher topology corrects are included.

It is worth to note that in the case where $\Gs =1$ and one can perform the summation to all orders
in $\Gs$, one finds that there is no higher order corrections to the $\tp^{-4}$ term in agreement 
with the expansion \rf{a3} to 3rd order on $\Gs$. Also, in the exact summation when $\Gs =1$ one 
finds that the coefficient to the term $\tp^{-7}$ is 4. This is in agreement with the expansion 
\rf{a3}-\rf{a4} provided all corrections of order four and higher to the  $\tp^{-7}$-term is zero.

For $\Gs = -4$ the 3rd order correction seems unreliable, as mentioned in the main text, 
but using the Pad\'e[2/1] approximant one can seemingly use much larger $|\Gs|$. In Fig.\ \ref{fig5}
we have shown calculation of $w_{\rm f}(z)$ for some range of $\Gs$ using the Pad\'e approximant.

\section*{Acknowledgment}
This research was supported in part by Perimeter Institute for Theoretical Physics. Research at Perimeter Institute is supported by the Government of Canada through the Department of Innovation, Science and Economic Development and by the Province of Ontario through the Ministry of Colleges and Universities. J.A. thanks the Perimeter Institute for hospitality while this work
was completed. The work was also supported by JSPS KAKENHI Grant Numbers JP25K07278.

\end{document}